\documentclass[a4paper,11pt]{article}
\usepackage{pos}
\usepackage{siunitx}
\usepackage{float}
\title{Tools and Advancements towards Data Standardization of the MAGIC Collaboration}

\author*[a]{Cyrus Walther}
\author[b]{Cosimo Nigro}
\author[a]{Dominik Elsässer}
\author[a]{Wolfgang Rhode}

\onbehalf{on behalf of the MAGIC Collaboration}

\affiliation[a]{Department of Physics, TU Dortmund University,\\
  Otto-Hahn-Straße 4a, 44227 Dortmund, Germany}

\affiliation[b]{Institut de Física d'Altes Energies, Universitat Autònoma Barcelona,\\
Edifici Cn, 08193 Bellaterra (Barcelona), Spain}

\emailAdd{cyrus.walther@tu-dortmund.de}

\abstract{

Gamma-ray astronomy is able to acquire large data volumes that astronomers
use to draw scientific conclusions from. Ensuring the possibility of accessing and
utilizing this data also after the lifetime of currently running
experiments requires the use of a standardized data format.
Following the data standardization format proposed by the gamma-ray astronomy
community, we present 104 h of the first production of 166 h of data from the MAGIC Imaging
Air Cherenkov Telescopes in standardized data format.
Six datasets were processed from which three are presented, all of which have
been analyzed and validated through comparison using the
open-source software Gammapy and the MAGIC analysis software MARS. \\

Furthermore, looking towards a large-scale production of standardized
data and a legacy of the data taken by the MAGIC experiment, we have developed and implemented the
automated database-driven MAGIC data reduction tool \textit{autoMAGIC}
which offers a reliable and reproducible way to produce high-level
datasets. By utilizing the automatization of parameter configuration choices,
the software allows for a reduction of human error as well as an acceleration in
the production of standardized data. Here, we also show comparable
results for data processed with manual and automatic methods.
}

\ConferenceLogo{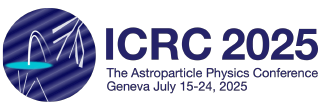}

\FullConference{39th International Cosmic Ray Conference (ICRC2025)\\
 15–24 July 2025\\
Geneva, Switzerland\\}

\begin{document}
\maketitle

\section{Introduction}

The scientific community of very-high-energy (VHE) gamma-ray astronomy has historically 
performed observations and analysis of data following an in-collaboration approach. This facilitated the development of customized tools tailored to the requirements and capabilities of the telescope. Parallel to the utilization of this practice, the exploration of analyses in the multi-wavelength (MWL) and multi-messenger regime on the basis of collaboration with other astronomy communities yields scientific gain, such as the development of a multi-messenger view of the Milky Way.\cite{ICECUBE} Following the motivation to tap the full potential of MWL studies and the next-generation gamma-ray telescope, the Cherenkov Telescope Array Observatory (CTAO), being designed as an open observatory, a demand for data standardization methods emerged within the VHE gamma-ray community. On these grounds, the Data Formats for Gamma-ray Astronomy (GADF) initiative was founded, initiating the discussion on advancing the development of specifications on data formats for high-energy gamma-ray data. \\

While the development of data standardization methods and techniques is intrinsic to the development of the CTAO, third-generation Imaging Air Cherenkov Telescopes, like the Major Atmospheric Gamma-ray Imaging Cherenkov (MAGIC) Telescopes, 
have already performed data acquisition for over two decades utilizing tailored solutions for data analysis. Concurrently, the large data fund gathered represents a potential of scientific endeavor. However, this potential extends above the capacities of the collaboration, offering the opportunity to yield scientific merit by opening access to these funds. However, due to tailored data analysis pipelines, this calls for the development and adoption of data standardization methods. Consequently, this striving for standardized data formats requires the development of an open-source analysis tool linking existing pipelines to new standardized formats as well as analysis tools building upon the new standardized data formats, such as Gammapy, that enable the creation of high-level physics results such as spectra and lightcurves.\cite{gammapy} \\

Moreover, the perspective of third-generation IACTs finalizing their time of service in the near future guides thoughts toward the long-term value of those experiments to the scientific community after their retirement. One aspect of the experiment's legacy value is represented by the data fund of the telescopes, which in parallel exceeds existing capacities for long-term data curation without data compression. Such a compressed legacy dataset likewise raises interest in a standardized data format, ensuring interoperability with data structures of emerging telescopes and compliance with the FAIR principles.\cite{FAIR} In addition, the data analysis and reduction necessary for the development of a compressed legacy dataset, requires a data reduction tool that holds the capacity to perform a consistent large-scale analysis of the data fund while concisely providing the metadata allowing traceability of the configuration of the steps performed during the analysis. \\

We present the development of methods producing the first $\SI{166}{\hour}$ MAGIC standardized data set, utilizing the magic\_dl3 tool, presenting different scientific analysis cases demonstrating the efforts towards the adoption of GADF standardized data criteria. In addition, we demonstrate the validation of these methods against existing data analysis methods performed in the MAGIC collaboration. 
Furthermore, we present the development of the database-driven tool \textit{autoMAGIC} capable of the consistent analysis of large data volumes while ensuring traceability as well as reproducibility. Comparably, we illustrate the validation of \textit{autoMAGIC} on a data set representing a variety of moonlight intensities.

\section{Development of standardized MAGIC Data}\label{sec:2}

The perspective of standardized data format guidelines has been shaped by the GADF recommendations with the resulting concept of data level 3 (DL3).\cite{GADF} DL3 is expected to include detector- and calibration-independent data which are stored in the GADF-compliant Flexible Image Transport System (FITS) file format, allowing for a stand-alone analysis towards higher-level scientific products.\cite{FITS} 
DL3 consists of two elements, an event-list of gamma-ray events as well as the parametrization of the systems response, the so-called instrument response function. Utilizing these two contents, detector-specific information such as event counts can be converted to experiment-independent units such as the flux of gamma rays. 

\begin{figure}[h]
    \centering
    \includegraphics[width=\linewidth]{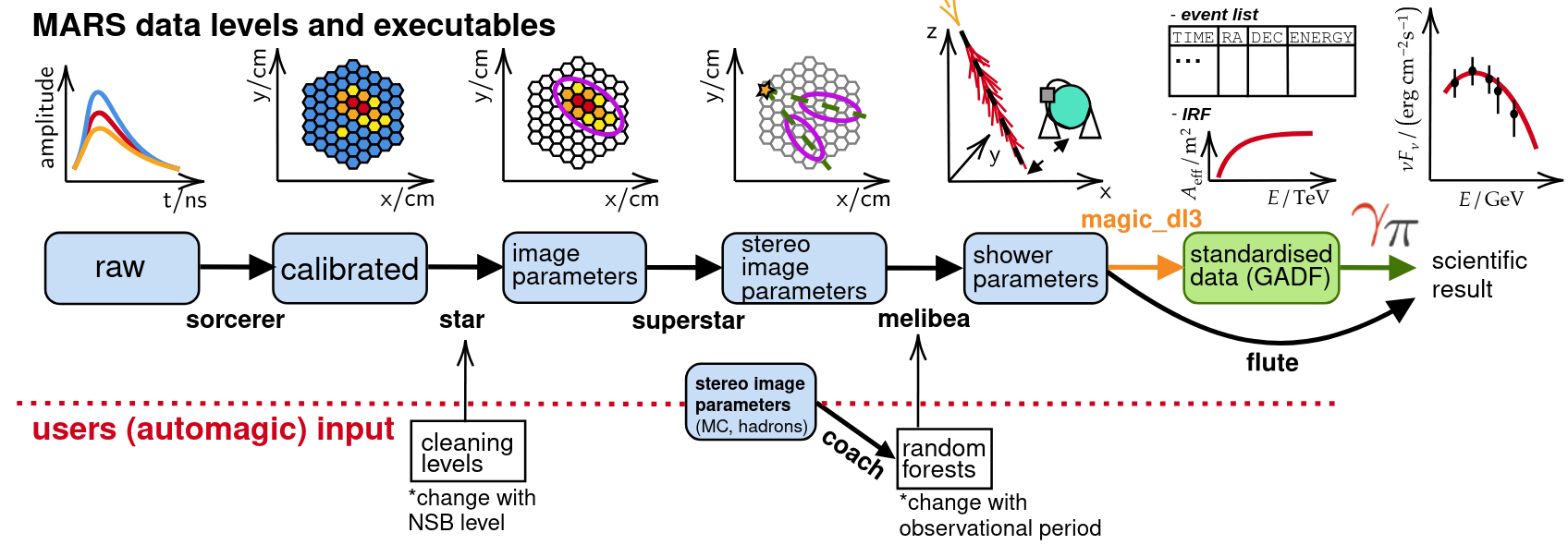}
    \caption{A structural float chart of the MAGIC data levels illustrating the visualized data in each step of the data analysis. Blue boxes describe the levels of data analysis with \textit{magic\_dl3} highlighted as the tool to produce standardized data following the GADF guidelines. Below the blue boxes, bold black terms describe MAGIC subprogram executables performed utilizing MARS. White boxes present the configuration input that is given to \textit{autoMAGIC} specifying the analysis characteristics.\cite{DL3}}
    \label{fig:DL3}
\end{figure}

Following Figure \ref{fig:DL3}, the chain of data analysis steps in MAGIC is illustrated. In that, the last blue box "shower parameters" represents the final
data analysis step before scientific results are obtained utilizing the program \textit{flute} in the MAGIC structure. Since the
GADF-compliant DL3 data level acts as a standardized data format from which scientific results can be acquired, it is necessary to develop a dedicated method that 
sources information from the "shower parameters" level and provides GADF-complained standardized data.  \\

Since none of the data-reduction steps performed by the MAGIC proprietary analysis software \textit{MARS} \cite{mars} result in a GADF-compliant data format, the tool \textit{magic\_dl3} was developed. The \textit{magic\_dl3} tool is a C++ analysis library that is built on \textit{ROOT} \cite{root}, \textit{MARS}, and \textit{CFITSIO} \cite{CFITSIO} providing MAGIC DL3 data on the basis of the proprietary \textit{MARS} melibea data level and converting them towards the FITS file format. \textit{magic\_dl3} obtains the first component of the information required for DL3 by applying a cut on the "gammaness" score resulting from the appliance of a trained random forest to the event information. The second component, the IRF, is obtained through building histograms that describe the collection area as well as the probability distribution from Monte Carlo simulations following the appliance of cuts identical to the performed data cuts. 

\section{autoMAGIC}

In order to address the challenge of performing the analysis steps up to the melibea data level shown in Figure \ref{fig:DL3} that are necessary for the basis of the \textit{magic\_dl3} library and to tackle the task of producing a consistent legacy data set, the database-driven automation tool \textit{autoMAGIC} has been developed. \textit{autoMAGIC} is able to adapt analysis configurations dependent on the respective observation conditions utilizing tables that connect the observation conditions with predefined configurations. Therefore, work-intensive manual adjustments of configurations for conditions such as sky brightness, atmospheric condition, or weather throughout the whole analysis time frame are automatized, otherwise resulting in high numbers of adjustable parameters required for performing a MARS data analysis. This effaces the possible human errors in the choice of parameters and allows for a more time-efficient analysis.

\begin{figure}[H]
    \centering
    \includegraphics[width=.6\linewidth]{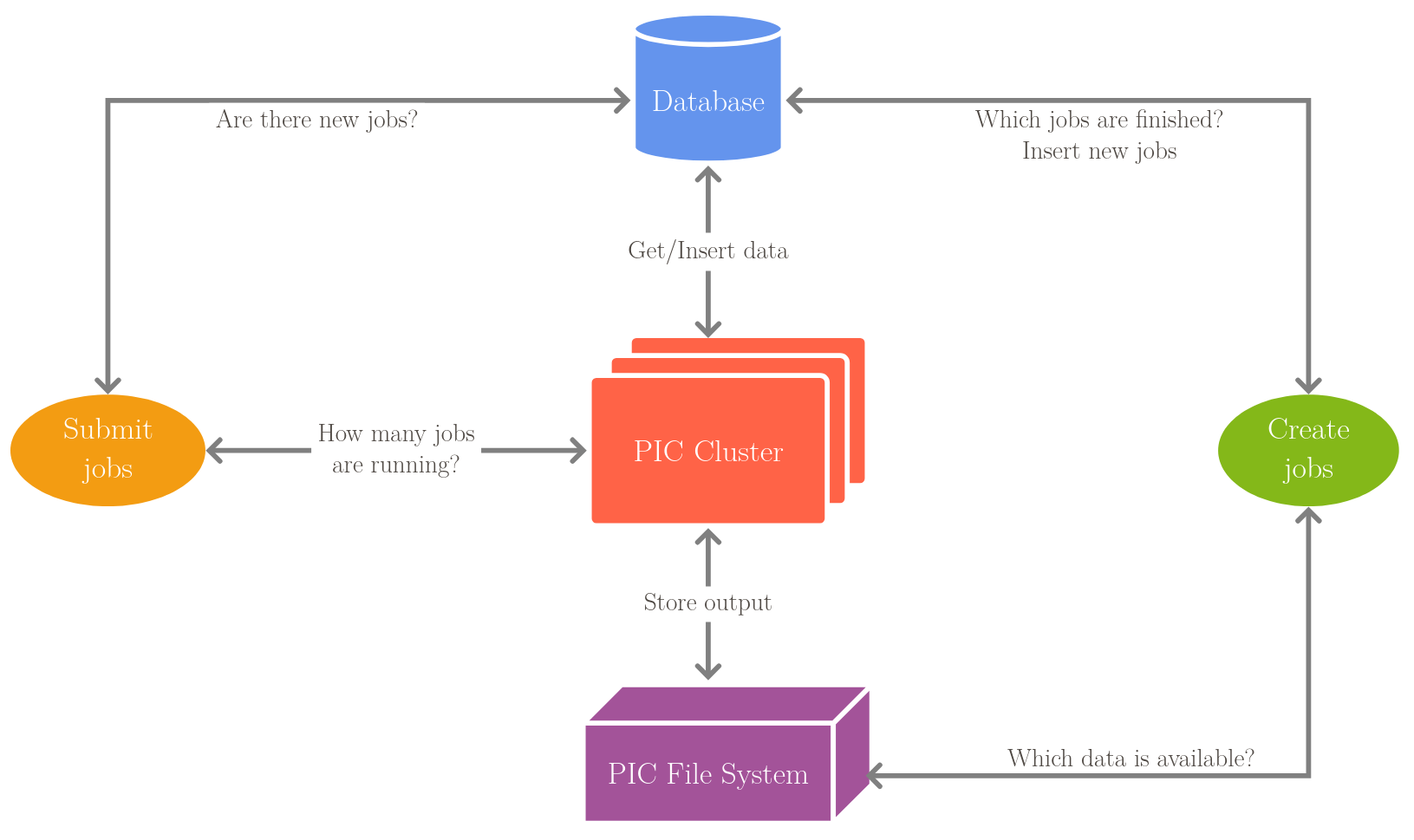}
    \caption{Description of the \textit{autoMAGIC} operation procedure. Data that characterizes the analysis is inserted into the database, which stores respective information, by requesting it from the PIC cluster. The PIC cluster draws the information from the PIC File System and returns it to the database. Jobs are created and inserted into the database based on the available data in the PIC File System and the existence of jobs in the database mirroring the analysis configuration of requested jobs. Jobs are submitted based on their status in the database and based on the workload of the PIC Cluster. \cite{JL}}
    \label{fig:aM}
\end{figure}

In the process, \textit{autoMAGIC} performs the choice of simulation data as well as background data following criteria for their source composition while scanning a multi-year data set, which is then used for the training of random forests. Subsequently, \textit{autoMAGIC} trains these random forests and applies them to obtain the information level of melibea data. To perform these steps, \textit{autoMAGIC} solely requires one input configuration featuring information such as the source name, the zenith range, the time range, the weather, and the sky brightness. Based on that information and subsequent information derived from the \textit{autoMAGIC} database tables, \textit{autoMAGIC} automatizes the creation of configuration files for executables of the \textit{MARS} software. Following the illustration of Figure \ref{fig:aM}, \textit{autoMAGIC} runs its computing tasks at the Port d'Informació Cientifica (PIC). Thus, \textit{autoMAGIC} exploits the advantage of precreation of the configuration files while utilizing the capacities of the PIC and its ability to parallel compute large amounts of analysis steps. Such analysis steps for which all necessary information has been computed will be, if computing resources allow, submitted to the cluster. Hence, the submission of jobs is always automatically performed at the earliest possible moment, allowing for significant savings in the required time.

\section{Method Validation}

\begin{figure}[H]
    \centering
    \includegraphics[width=\linewidth]{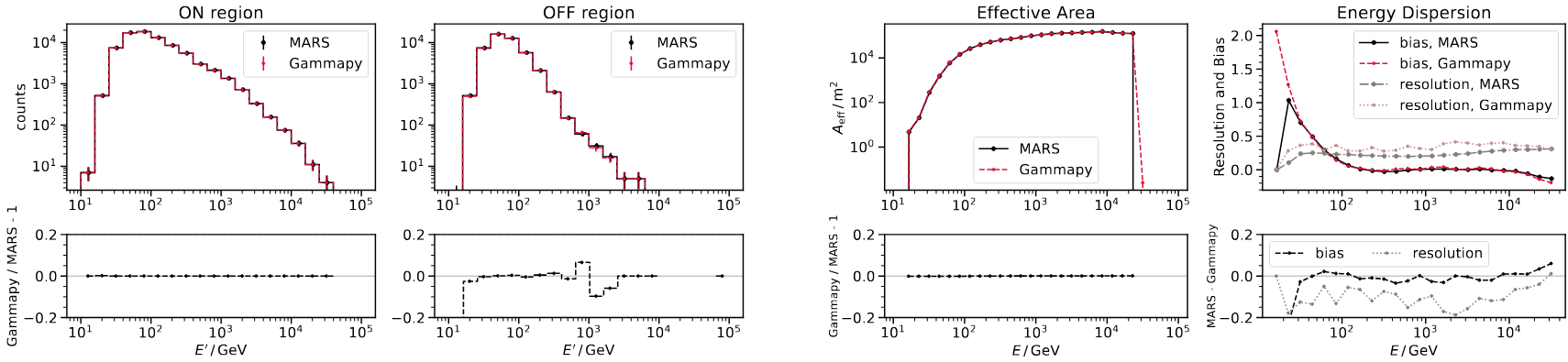}
    \caption{Comparison of the DL3 components as outlined in section \ref{sec:2}. The most left plot depicts the histogram of counts in the on-region with the second left plot depicting the histogram of counts in the off-region. The most right plot depicts the bias and the resolution of the energy dispersion while the second right plot depicts the effective area. All plots include a ratio demonstrating the dispersion of both methods.\cite{DL3}}
    \label{fig:gammapy_validation}
\end{figure}

To validate the \textit{magic\_dl3} and the \textit{autoMAGIC} tool, different data samples were chosen, addressing different aspects of gamma-ray data analysis and allowing for a thorough validation of the presented tools. Firstly, the number of counts with respect to the events' energy is validated,  visualizing deviations of the shape of counts between data analyzed with the \textit{MARS} pipeline and the pipeline utilizing \textit{magic\_dl3} together with \textit{Gammapy}. Similarly, the effective area and the energy dispersion are examined to validate the IRF both methods provide. This approach is shown for the on and off region for a $\SI{0.4}{\degree}$ single-offset $\SI{30}{\hour}$ Crab Nebula sample. Figure \ref{fig:gammapy_validation} shows these four comparisons and agreement can be found on all plots. It is to be considered that the last data point of the effective area is due to a difference in the \textit{Gammapy} and \textit{MARS} interpolation methods.

In a second validation step, the impact of wobble-offsets \cite{wobble} is analyzed, addressing the aspect of validating the observation of diffuse sources. Seven different wobble-offsets have been chosen with a total of $\SI{42}{\hour}$ of Crab Nebula data. A set of $\SI{30}{\hour}$ and [5°, 35°] zenith range with a wobble-offset of $\SI{0.4}{\degree}$ is chosen and six sets of $\SI{20}{\hour}$ and [5°, 50°] are chosen with wobble-offsets of: $\SI{0.2}{\degree}$, $\SI{0.35}{\degree}$, $\SI{0.4}{\degree}$, $\SI{0.7}{\degree}$, $\SI{1.0}{\degree}$, $\SI{1.4}{\degree}$. For the spectrum, we assume a log-parabola spectral model with $E_0=\SI{1}{\tera\electronvolt}$. 

\begin{equation}
    \frac{d\phi}{dE}(E;\phi_0, \alpha, \beta, E_0)=\phi_0 \Bigl(\frac{E}{E_0}\Bigl)^{-\alpha -\beta \log_{10}(\frac{E}{E_0})}
    \label{eq}
\end{equation}

\begin{figure}[H]
    \centering
    \includegraphics[width=.75\linewidth]{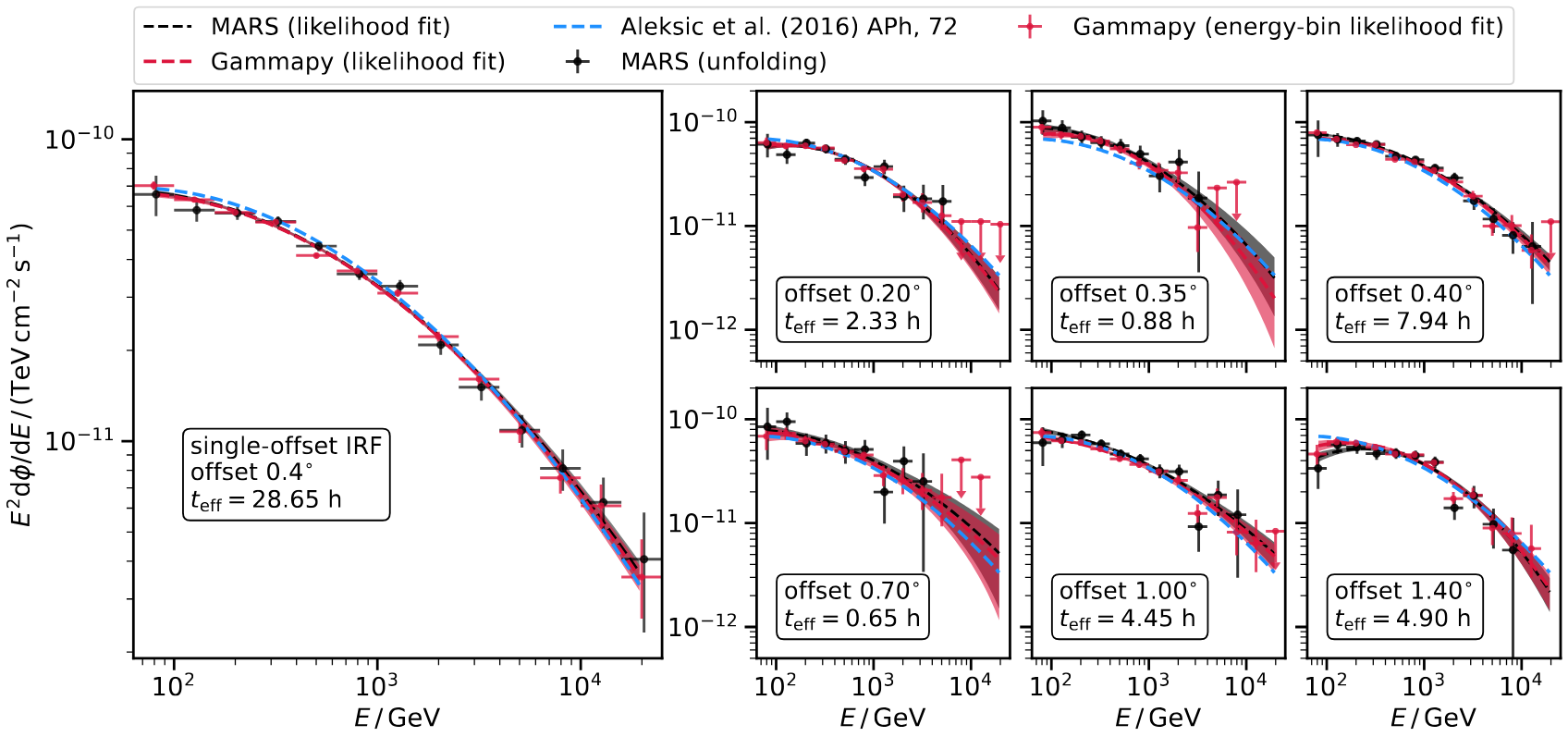}
    \caption{Obtained SEDs of the Crab Nebula with the left-side plot showing the SED of a $\SI{0.4}{\degree}$ wobble-offset. The six subplots on the right side depict the SED depending on the wobble-offset ranging from $\SI{0.2}{\degree}$ in the top-left to  $\SI{1.4}{\degree}$ in the bottom right. In all plots, red describes the novel pipeline using \textit{magic\_dl3} and \textit{Gammapy}, black using the proprietary \textit{MARS} approach, and blue showing the reference value.\cite{DL3}}
    \label{fig:gammapy_SED}
\end{figure}

\begin{figure}[H]
    \centering
    \includegraphics[width=.75\linewidth]{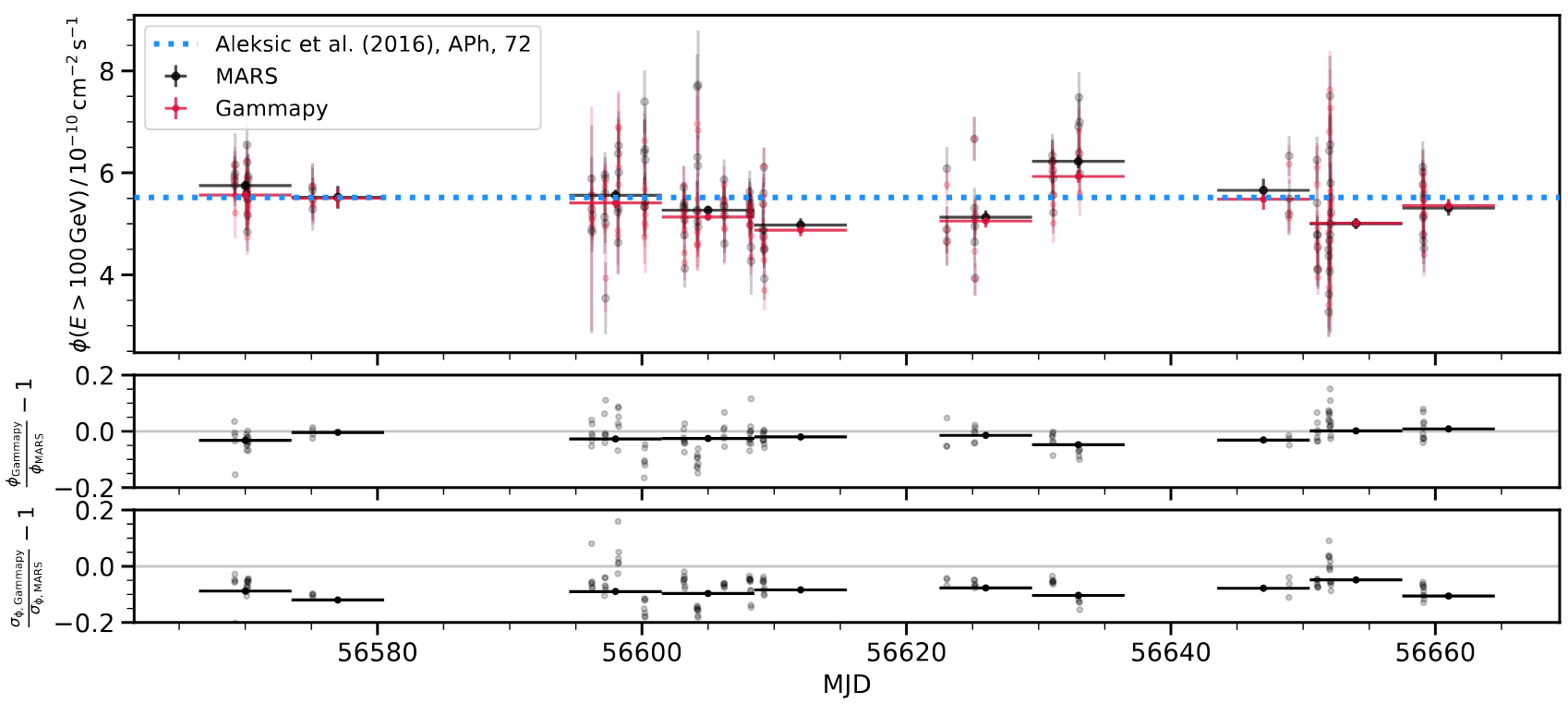}
    \caption{The light curve over the whole $\SI{42}{\hour}$ data set. Therein, the run-wise light curve is shown in transparent points and the weekly lightcurve is shown in solid points. The blue dots represent the reference from MAGIC. \cite{aleksic} \cite{DL3}}
    \label{fig:gammapy_LC}
\end{figure}

Following the log-likelihood fit of \ref{eq}, we obtain the spectra depicted in Figure \ref{fig:gammapy_SED}. We observe good agreement of both pipelines in all offset subplots. Furthermore, we obtain a light curve in Figure \ref{fig:gammapy_LC} of the overall dataset. Both binnings, run- and week-wise, show good agreement, examining deviations illustrated in ratio plots of less than $\SI{5}{\percent}$ for the week-wise binning and less than $\SI{20}{\percent}$ for the night-wise binning.

\begin{figure}[H]
    \centering
    \includegraphics[width=.75\linewidth]{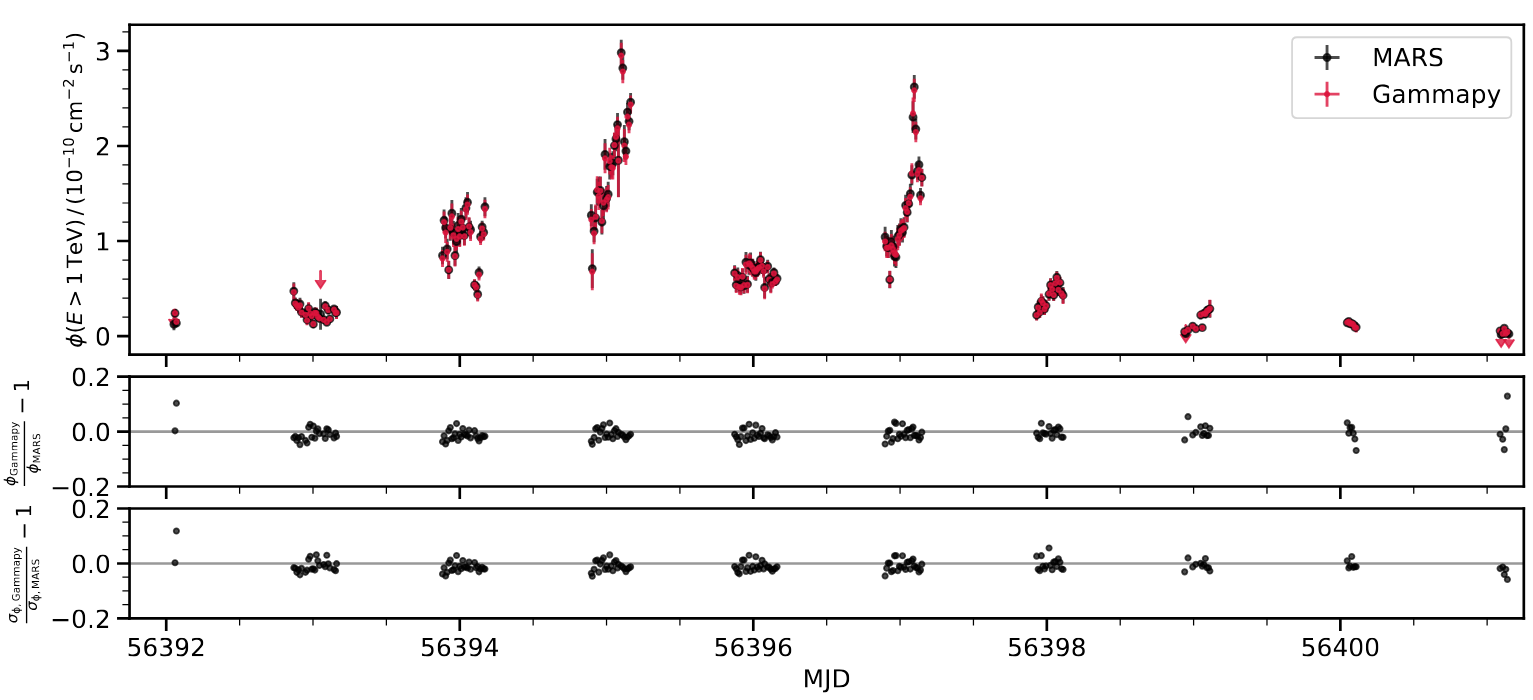}
    \caption{Depiction of the lightcurve of $\SI{42}{\hour}$ of Mrk421 data, showing the calculated run-wise values for both analysis pipelines. Ratios of the flux and the signal indicate low deviations. \cite{DL3}}
    \label{fig:Mrk421-LC}
\end{figure}

Thirdly, testing the analysis of the flux of a highly variable source, a dataset of $\SI{42}{\hour}$ from 2014 of Mrk421 in the zenith range of [5°, 70°] has been chosen. This data has already been analyzed in \cite{mrk paper}, allowing for comparison with the manual method. The light curve presented in Figure \ref{fig:Mrk421-LC} shows good agreement between the proprietary MAGIC software and the DL3 software with the ratio of the flux never exceeding a deviation of 0.2.

\begin{figure}[H]
    \centering
    \includegraphics[width=.75\linewidth]{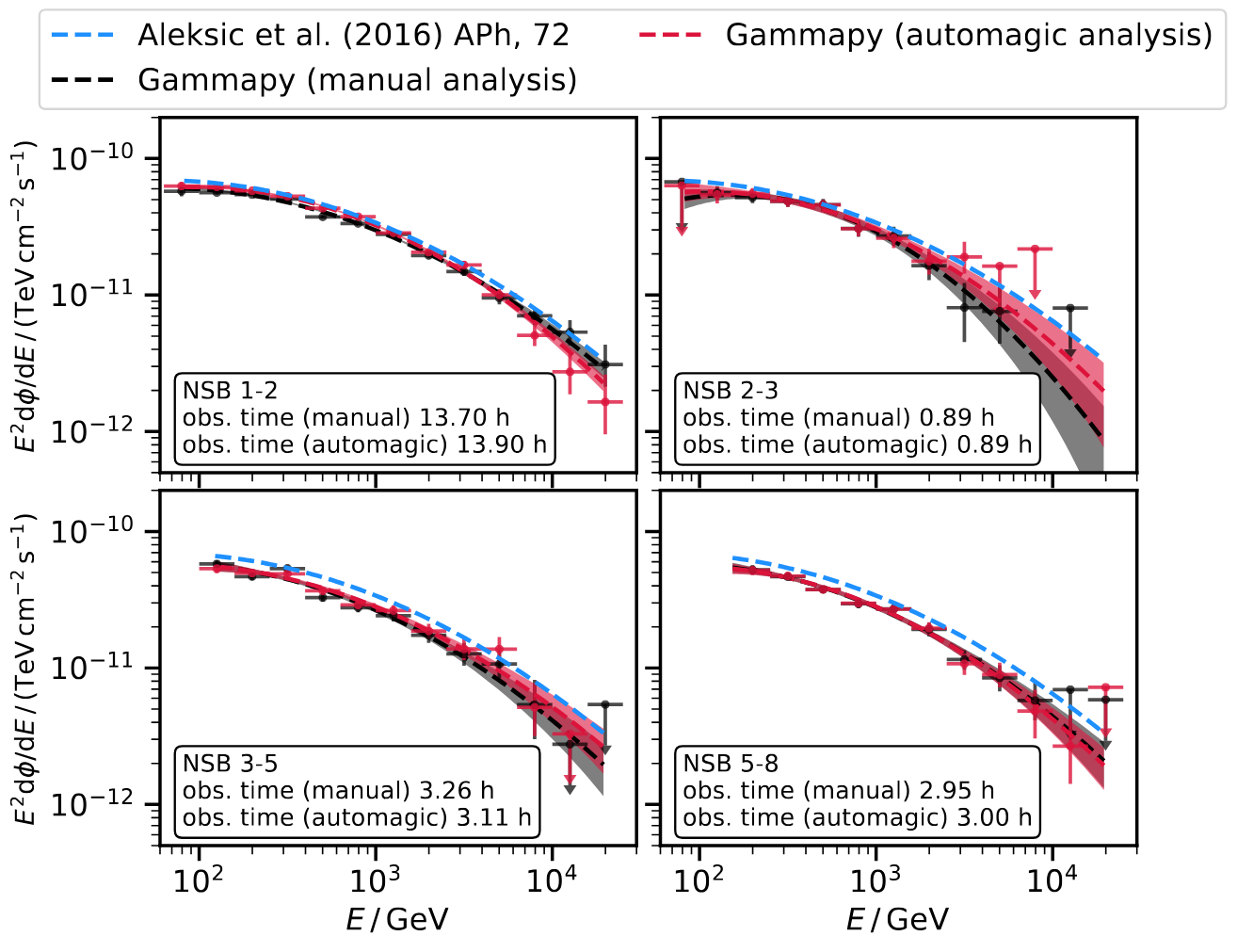}
    \caption{Illustration of the spectra of Crab Nebula data obtained with the manual method, marked in black, and the \textit{autoMAGIC} method, marked in red. A reference to the analyzed data is shown in blue.\cite{aleksic} The top-left plot depicts the spectrum with the lowest NSB bin. The top-right plot shows the spectra calculated on data of the second lowest NSB bin with the bottom-left and the bottom-right showing the third and fourth lowest NSB bin. \cite{DL3}}
    \label{fig:autoMAGIC_results}
\end{figure}

The validation of \textit{autoMAGIC} requires addressing those aspects of \textit{autoMAGIC} that differ from manual analysis. Hence, a dataset of $\SI{20}{\hour}$ of varying moonlight conditions of the Crab Nebula between 11-2018 and 09-2019 in a zenith range of [5°, 50°] was chosen, addressing the automation of performing moonlight analysis specifications appropriately. For this validation, the high-level analysis of the manual and the \textit{autoMAGIC} dataset have both been performed on DL3 data utilizing Gammapy. Following \cite{moon}, the data has been divided into four bins of night sky background (NSB) rates. Figure \ref{fig:autoMAGIC_results} shows the spectra for both methods in the four background bins. All four bins show similar observation times with discrepancies of less than $\SI{5}{\percent}$. We observe good agreement between the spectra obtained through the usage of \textit{autoMAGIC} and the spectra obtained manually. Moreover, in the two higher bins, NSB 3-5 and NSB 5-8, an overall underestimation of the Crab Nebula data is observed, agreeing with a dedicated study of moon impacts. \cite{moon}

\section{Summary}

We presented the first production of MAGIC data in a GADF-compliant and standardized format, demonstrating the efforts to convert to standardized data formats and \textit{magic\_dl3} as the MAGIC library facilitating this change. Furthermore, we address the need for a consistent tool for the automatized and reliable analysis of MAGIC data by presenting \textit{autoMAGIC}. We validated the methods utilizing three different datasets for different scientific purposes and received overall good results, comparing \textit{autoMAGIC} as well as \textit{magic\_dl3} with \textit{Gammapy} with the proprietary MAGIC software \textit{MARS}.

\section*{Acknowledgements}
\scriptsize
We would like to thank the Instituto de Astrof\'{\i}sica de Canarias for the excellent working conditions at the Observatorio del Roque de los Muchachos in La Palma. The financial support of the German BMBF, MPG and HGF; the Italian INFN and INAF; the Swiss National Fund SNF; the grants PID2019-107988GB-C22, PID2022-136828NB-C41, PID2022-137810NB-C22, PID2022-138172NB-C41, PID2022-138172NB-C42, PID2022-138172NB-C43, PID2022-139117NB-C41, PID2022-139117NB-C42, PID2022-139117NB-C43, PID2022-139117NB-C44, CNS2023-144504 funded by the Spanish MCIN/AEI/ 10.13039/501100011033 and "ERDF A way of making Europe; the Indian Department of Atomic Energy; the Japanese ICRR, the University of Tokyo, JSPS, and MEXT; the Bulgarian Ministry of Education and Science, National RI Roadmap Project DO1-400/18.12.2020 and the Academy of Finland grant nr. 320045 is gratefully acknowledged. This work was also been supported by Centros de Excelencia ``Severo Ochoa'' y Unidades ``Mar\'{\i}a de Maeztu'' program of the Spanish MCIN/AEI/ 10.13039/501100011033 (CEX2019-000920-S, CEX2019-000918-M, CEX2021-001131-S) and by the CERCA institution and grants 2021SGR00426 and 2021SGR00773 of the Generalitat de Catalunya; by the Croatian Science Foundation (HrZZ) Project IP-2022-10-4595 and the University of Rijeka Project uniri-prirod-18-48; by the Deutsche Forschungsgemeinschaft (SFB1491) and by the Lamarr-Institute for Machine Learning and Artificial Intelligence; by the Polish Ministry Of Education and Science grant No. 2021/WK/08; and by the Brazilian MCTIC, CNPq and FAPERJ.

\end{document}